\documentclass[showpacs, amsmath,amssymb, aps,showkeys]{revtex4}
\usepackage{amsmath,amssymb,amsfonts,latexsym,float,graphics,epsfig}
\def\R{\mathbb{R}}
\def\T{\mathbb{T}}

\catcode`@=11 \@addtoreset{equation}{section}

\newcommand{\be}{\begin{equation}}
\newcommand{\ee}{\end{equation}}

\begin{document}
\title{Emergence of the Gambier equation in cosmology}
\author{D. Batic}
\email{davide.batic@ku.ac.ae}
\affiliation{%
Department of Mathematics,\\  Khalifa University of Science and Technology,\\ Main Campus, Abu Dhabi,\\ United Arab Emirates}
\author{A. Ghose Choudhury}
\email{aghosechoudhury@gmail.com}
\affiliation{
Department of Physics \\ Diamond Harbour Women's University\\ 
D.H. Road, Sarisha 743368, West Bengal,  India
}
\author{Partha Guha}
\email{partha.guha@ku.ac.ae}
\affiliation{%
Department of Mathematics,\\  Khalifa University of Science and Technology,\\ Main Campus, Abu Dhabi,\\ United Arab Emirates}

\date{\today}
\begin{abstract}
We show how the Gambier equation arises in connection to Friedmann-Lema$\mbox{\^{i}}$tre-Robertson-Walker (FLRW) cosmology and a Dark Matter equation of state. Moreover, we provide a correspondence between the Friedmann equations and the Gambier equations that possess the Painlev$\acute{\mbox{e}}$ property in $2+1$ dimensions. We also consider special cases of the Gambier G27 equation such as the generalized Pinney equation. For an extended FLRW model with dynamic scalar field as matter model, the Einstein equations correspond to the Milne-Pinney equation which in turn can be mapped to the parametric Gambier equation of second order.
\end{abstract}


\maketitle



\section{Introduction}
Since Witten' seminal paper in 1988 \cite{Witten}, where he showed that $2+1$-gravity is an integrable system, there has been a continuous interest on this topic in the scientific community. Even though three-dimensional gravity exhibits some serious shortcomings such as the absence of gravitational waves, lack of a Newtonian limit and no dynamical degrees of freedom \cite{Giddings}, it represents a useful arena because despite its simplicity it displays non-trivial features such as singularities \cite{Barrow} and black holes \cite{Tit1,Tit2}. Moreover, toy models formulated in $2+1$-gravity may be also relevant to four-dimensional gravity, its quantization problem and cosmology. It was already observed by Giddings et al.  \cite{Giddings} that such models can be used to probe into quantization in the presence of gravity because one can add an external field with its own degrees of freedom that can be quantized. In addition, it was shown in \cite{Giddings} that the solution of the Friedmann equations for a radiation-dominated universe in three-dimensions behaves like that for dust in one higher dimension. It is worth to mention along this line that Barrow et al. \cite{Barrow} proved that certain static and non-static solutions of $2+1$-gravity generated by a source modelled in terms of a scalar field, perfect fluid or a magnetic field exhibit a correspondence with the usual $3+1$-theory of gravitation by means of a Kaluza-Klein type reduction. He also showed therein that $2+1$-cosmological models involving self-interacting scalar fields can be relevant to probe into the production mechanisms of inflationary behaviour. The mantra emerging from these examples is that results obtained in $2+1$-gravity should not be dismissed on physical grounds because they may have a history in the standard theory of general relativity. For this reason, we will consider cosmologies in $3+1$-dimensions as well as in $2+1$-dimensions. 

The study of Einstein's field equations in scalar field cosmologies are essentially motivated from high energy physics and quantum gravity theories although the observational discovery of the field is still lacking \cite{Barrow,Cornish,Cruz,CGHKPW,WK,Krit}. Such cosmologies (both for the isotropic and anisotropic models) may be mapped to equations of the generalized Ermakov-Milne-Pinney (EMP) type \cite{CGHKPW,WK,HL,Lidsey,Cariglia}. Hawkins and Lidsey \cite{HL} developed an analytical approach to models of this type by expressing the cosmological field equations in terms of an Ermakov-Pinney system, such system often arises in studies of nonlinear optics, nonlinear elasticity, fluid dynamics, Bose-Enstein condensation and various other areas in physics (For further references, see, e.g. \cite{Leach}). In general, an Ermakov system is a pair of coupled, second-order, nonlinear ordinary differential equations (ODEs) but in the one-dimensional case, the two equations decouple and the system reduces to a single equation known as the Ermakov-Pinney equation \cite{Erma,Pinney}
\be\label{EP} \ddot{x}+\omega^2(t) x=\frac{1}{x^3}.\ee
The solution of this equation is of the form
\be x(t)=\sqrt{Au^2+2Buv+Cv^2}\ee where $u$ and $v$ are any two linearly independent solutions of
\be\label{a3} \ddot{x}+\omega^2(t) x=0\ee and the constants $A,B$ and $C$ are related by $B^2-AC=1/W^2$,
where $W$ is the constant Wronskian of the two linearly independent solutions of (\ref{a3}). The Ermakov-Pinney equation
 appears in diverse areas of physics such as nonlinear optics, hydrodynamics and has been extensively studied in the
context of FLRW cosmology  \cite{Lidsey,HK}.  Pinney's fundamental contribution consisted
in presenting the solution of (\ref{EP}) as a first integral of the third-order equation
$$\dddot{x}+4\omega^2 \dot{x}+4\omega\dot{\omega} x=0.$$
The general properties of (\ref{EP}) together with its invariants are well known \cite{Guenther,LL}.

In the early years of the twentieth century Gambier listed a minimal set of 24 equations of the
second-order which possessed the Painlev\'{e} property \cite{Gam,Ince}. The Gambier equations have however
found comparatively little application in the applied sciences, though equation G15 of Gambier's list \cite{Gam}
\begin{equation}\label{GXV}
Y^{''}=\frac{(Y^{'})^2}{Y}+\frac{Y^{'}}{Y}+\xi(X)Y^2-Y\frac{d}{dX}\left(\frac{\xi^{'}}{\xi}\right)
\end{equation}
with $\xi$ some differentiable function is known to be related to the family of generalized Ermakov-Pinney equations. In special cases, this can be
transformed into a Hamiltonian system and at times even into a conjugate Hamiltonian system which
may be solved explicitly \cite{GGCG,GGC}.

An alternative (non-EMP) formulation of a
homogeneous, isotropic scalar field cosmology has been studied in
\cite{WK,DW}. In this case the mapping between the FLRW cosmology \cite{Weinberg} and the Ermakov-Pinney fails. The same is also true when
the scalar field is time-dependent. The purpose of the present article  is to explore the possibility of applications
of the equations of Gambier's classification in the domain of FLRW cosmology. The Gambier equation which
we obtain from the reduction of $2+1$-FLRW cosmology is hyperbolic in nature, as a result all the reductions of the
cosmological Gambier equation are hyperbolic in nature. If we take the scalar field of the Williams-Kevrekidis model to be time-dependent, then the reduced  $2+1$-FLRW cosmology contains a time-dependent drag term. The simplified model of this cosmological equation is the Milne-Pinney equation. In this context, we show that the generalized FLRW equation is a reduction of the time-dependent extension of the Gambier equation G15. A brief note on this time-dependent extension of equation G15  which is also sometimes referred to as the parametric Gambier equation is given in an appendix. Furthermore we also indicate how equations of the Ermakov-Painlev\'{e} II type arise in the extended FLRW cosmological equation with a scalar field and perfect fluid matter source. The extended FLRW cosmological equation with such a scalar field and perfect fluid matter has been the object of previous analysis by \cite{DW}.

In recent times, a hybrid Ermakov-Painlev\'{e} II system was derived by  Rogers \cite{Rogers} in a pioneering work as a
reduction of a coupled $N+1$-dimensional
Manakov-type nonlinear Schr\"{o}dinger system. He showed that the Ermakov invariants admitted by the hybrid system were key
to its systematic reduction in terms of a single component Ermakov-Painlev\'e II equation which, in
turn, may be linked to the integrable Painlev\'e II equation. Rogers \cite{Rogers1} further extended the hybrid family
to derive Ermakov-Painlev\'e IV equation
and investigated the  connection between the FLRW cosmology of \cite{DW} and  the generalized Pinney equations derived by
Rogers, Schief and Winternitz \cite{RSW}. We may
exploit the underlying Ermakov type invariants of these equations to obtain solutions of the reduced
FLRW cosmological equation with scalar field and perfect fluid matter source. In summary therefore, the chief objective of the present paper is to bridge the  connection between various Gambier equations, generalized Pinney equations and  the extended FLRW cosmological equation.
In particular, we develop an analytical approach to generalized FLRW models  by expressing the cosmological field equations in terms of the Gambier equations and generalized Pinney equations.

\section{$3+1$-FLRW cosmology and the linearizable Gambier equations}

As mentioned in the introduction a possible arena  in which the equations of Gambier's
classification can find applications is in the domain of FLRW
cosmology. Several papers have been devoted to the issue of various
ordinary differential equations appearing in FLRW cosmologies
\cite{Lidsey,CGHKPW,HL,HK,Myrzakulov}. In such cosmologies one starts with the standard
gravitational action
\be\label{action} S=\frac{1}{16\pi}\int(R+L_m)\sqrt{-g}\;d^4x,\ee
where $R$ is the scalar curvature of the space-time described by the metric tensor $g_{\mu\nu}$ with signature $(-,+,+,+)$ and $g=\mbox{det}(g_{\mu\nu})$, $L_m$  represents 
the Lagrangian in the matter sector with the FLRW
space-time being characterized by the scale factor $a$ and the metric in geometrized units ($c=\hbar=G=1$)
\be\label{met1}
ds^2=-dt^2+a^2(t)\left(\frac{dr^2}{1-kr^2}+r^2 d\vartheta^2 + r^2\sin^2{\vartheta} d\varphi^2\right).
\ee
Here, $k=-1,0,1$ describe a manifold with constant negative curvature (open universe), no curvature (flat universe)
and positive curvature (closed universe), respectively. Out of the $16$ coefficients of the metric tensor 
only four are non-vanishing in the FLRW metric
and they include only one unknown function, the
scaling factor $a$. After substituting these expressions into the Einstein equations 
\begin{equation}
R_{\mu\nu}-\frac{1}{2}g_{\mu\nu}R=8\pi T_{\mu\nu},
\end{equation}
where $R_{\mu\nu}$ denotes the Ricci tensor and $T^{\mu\nu}=\mbox{diag}(\rho,p,p,p)$ is the energy-momentum tensor of an isotropic fluid with energy density $\rho$ and pressure $p$, it turns out that only two equations are independent, namely
\be\label{XX}
\left(\frac{\dot{a}}{a} \right)^2 + \frac{k}{a^2} = \frac{8}{3}\pi \rho, \qquad
2\frac{\ddot{a}}{a} + \left(\frac{\dot{a}}{a} \right)^2 + \frac{k}{a^2} = - {8\pi}p,
\ee
where $\rho$ is the mass-energy density and $p$ is the pressure. The first equation
is known as the Friedmann equation.
If we subtract the first equation from the second one, we obtain the so-called acceleration equation, and from these
equations one may obtain the conservation equation
\be
\label{E3}\dot{\rho}+3H(\rho+p)=0,\ee where $H=\dot{a}/a$ is the
Hubble parameter. The overdot here denotes differentiation with
respect to $t$. It is not difficult to verify that the equations (\ref{XX}) and (\ref{E3}) can be expressed in terms of $N=\ln{a}$ as 
\begin{eqnarray}
p&=&-\frac{1}{8\pi}\left(2\ddot{N}+3\dot{N}^2+ke^{-2N}\right),\label{E4}\\
\rho&=&\frac{3}{8\pi}\left(\dot{N}^2+ke^{-2N}\right),\label{E5}
\end{eqnarray}
while the conservation equation becomes
\be\label{E6}\dot{\rho}=-3\dot{N}(p+\rho).\ee 
{\bf{
It is noteworthy that the equations (\ref{XX}) and (\ref{E3}) exhibit a relationship. To illustrate this aspect, it can be observed that the second equation in (\ref{XX}) is redundant. This can be easily seen if we express the pressure $p$ using (\ref{XX}) as follows
\begin{equation}\label{palias}
p=-\frac{1}{8\pi}\left[2\frac{\ddot{a}}{a}+\left(\frac{\dot{a}}{a} \right)^2 + \frac{k}{a^2}\right].
\end{equation}
Furthermore, by isolating $p$ from (\ref{E3}), we can obtain
\begin{equation}\label{pbis}
p=-\frac{\dot{\rho}}{3H}-\rho.
\end{equation}
At this stage, we can solve the first equation in (\ref{XX}) for $\rho$, which enables us to compute $\dot{\rho}$ and replace $\rho$ and $\dot{\rho}$ into (\ref{palias}). A straightforward computation reveals that the right-hand side of (\ref{pbis}) is identical to that of (\ref{palias}).
}} 
In \cite{Myrzakulov,Dark}
the authors considered several FLRW cosmologies in connection with Dark Matter (DM) models in which the  equations of
state (EOS) are a special case of the more general EOS 
\begin{equation}\label{druk}
p=f_1(\dot{N}, N, t)\rho+f_2(\dot{N}, N, t)\rho^\beta+f_3(\dot{N}, N,
t),
\end{equation}
where $\beta\ne 1$ does not need to be an integer and $f_i$ with $i=1,2,3$ are some
functions of $\dot{N}, N$ and $t$. If we replace (\ref{druk}) into the conservation equation and use (\ref{E5}), we end up with a new master equation for $N$ given by 
\begin{equation}\label{master}
\ddot{N}=-\frac{3}{2}\left[1+f_1(\dot{N}, N, t)\right]\left(\dot{N}^2+ke^{-2N}\right)
-\frac{3^\beta f_2(\dot{N}, N, t)}{2^{3\beta-2}\pi^{\beta-1}}\left(\dot{N}^2+ke^{-2N}\right)^\beta
-4\pi f_3(\dot{N}, N, t)+ke^{-2N}.
\end{equation}
At this point we are free to choose the functions $f_i$ and the parameters $k$ and $\beta$ so that the above equation reduces to some known ODE. In particular, by setting $k=0$ in (\ref{master}) we {\bf{come up with the following cosmologies which have been partially discussed in \cite{Myrzakulov,Dark}.}}
\begin{itemize}
\item
{\bf{The Pinney's cosmology represented by G15 in (\ref{GXV}) with $Y=N$, $X=t$ and $^{'}$ replaced by $\dot{}=d/dt$ is a special case of (\ref{master}) with $k=0$, $\beta=1/2$ and}} 
\begin{equation}
f_1=-\frac{2+3N}{N},\quad
f_2=-\frac{1}{\sqrt{6\pi}N},\quad
f_3=-\frac{1}{4\pi}\left[\xi(t)N^2-N\frac{d}{dt}\left(\frac{\dot{\xi}}{\xi}\right)\right]
\end{equation}
{\bf{with $\xi$ some differentiable function. This is a generalization of the Pinney cosmology discussed in \cite{Myrzakulov,Dark}.}}
\item
{\bf{Schr\"{o}dinger's cosmology described by $\ddot{N}=\xi(t)N+\sigma N$ considered by \cite{Dark} is recovered by choosing}}
\begin{equation}
f_1=-1,\quad
f_2=0,\quad
f_3=-\frac{N}{4\pi}\left[\sigma+\xi(t)\right]
\end{equation}
{\bf{with $\xi$ some differentiable function and $\sigma\in\mathbb{R}$ a free parameter. Note that since $f_2$ vanishes, the parameter $\beta$ in (\ref{master}) does not need to be fixed.}}
\item
{\bf{Hypergeometric cosmology controlled by the differential equation \cite{Dark}}}
\begin{equation}
\ddot{N}=\frac{(a+b+1)t-c}{t(1-t)}\dot{N}+\frac{ab}{t(1-t)}N,\quad a,b,c\in\mathbb{R}
\end{equation}
{\bf{is a special case of (\ref{master}) with $\beta=1/2$}} 
\begin{equation}
f_1=-1,\quad
f_2=-\frac{(a+b+1)t-c}{\sqrt{6\pi}t(t-1)},\quad
f_3=-\frac{abN}{4\pi t(1-t)}.
\end{equation}
\item
{\bf{Painlev$\acute{\mbox{e}}$ cosmologies of the type (see corresponding Painlev$\acute{\mbox{e}}$ equations in \cite{Dark})}}
\begin{itemize}
\item
P1: {\bf{we have the following possible choices
\begin{equation}
f_1=-1,\quad f_2=0,\quad f_3=-\frac{6N^2+t}{4\pi},
\end{equation}
with any $\beta\in\mathbb{R}$ or 
\begin{equation}
\widehat{f}_1=-1,\quad\beta=0,\quad \widehat{f}_2+\widehat{f}_3=-\frac{6N^2+t}{4\pi}.
\end{equation}
}}
\item
P2: {\bf{we find
\begin{equation}
f_1=-1,\quad f_2=0,\quad f_3=-\frac{2N^3+(t-t_0)N+\alpha}{4\pi},
\end{equation}
with any $\beta\in\mathbb{R}$ or 
\begin{equation}
\widehat{f}_1=-1,\quad\beta=0,\quad \widehat{f}_2+\widehat{f}_3=-\frac{2N^3+(t-t_0)N+\alpha}{4\pi},\quad t_0\in\mathbb{R}.
\end{equation}
}}
\item
P3: {\bf{we have}}
\begin{equation}
f_1=-\left(1+\frac{2}{3N}\right),\quad\beta=\frac{1}{2},\quad f_2=\frac{1}{\sqrt{6\pi}t},\quad f_3=\frac{1}{4\pi t}\left(\widehat{\alpha}N^2+\widehat{\beta}\right)-\frac{\widehat{\gamma}}{4\pi}N^3-\frac{\widehat{\delta}}{4\pi N}
\end{equation}
with $\widehat{\alpha}$, $\widehat{\beta}$, $\widehat{\gamma}$ and $\widehat{\delta}$ real parameters entering in P3. Another more general choice is
\begin{equation}
\widetilde{f}_1+\widetilde{f}_2=-\left(1+\frac{2}{3N}\right),\quad\beta=1,\quad
\widetilde{f}_3=f_3.
\end{equation}
\item
P4: {\bf{we obtain}}
\begin{equation}
f_1+f_2=-\left(1+\frac{1}{3N}\right),\quad\beta=-1,\quad 
f_3=-\frac{1}{4\pi N}\left[\frac{3}{2}N^4+4tN^3+2(t-\widehat{\alpha})N^2+\widehat{\delta}\right].
\end{equation}
\item
P5: {\bf{if we consider $\beta=1/2$, then}}
\begin{equation}
f_1=-\left[1+\frac{1}{3N}+\frac{2}{3(N-1)}\right],\quad
f_2=\frac{1}{\sqrt{6\pi}t},\quad
f_3=-\frac{1}{4\pi}\left[\frac{\widehat{\gamma}}{t}N+\frac{(N-1)^2}{t^2}\left(\widehat{\alpha} N+\frac{\widehat{\beta}}{N}\right)+\widehat{\delta}\frac{N(N+1)}{N-1}\right].
\end{equation}
{\bf{If we instead pick $\beta=1$, we end up with}}
\begin{equation}
\widetilde{f}_1+\widetilde{f}_2=-\left[1+\frac{1}{3N}+\frac{2}{3(N-1)}\right],\quad
\widetilde{f}_3=-\frac{1}{4\pi}\left[-\frac{1}{t}\left(\dot{N}-\widehat{\gamma}N\right)+\frac{(N-1)^2}{t^2}\left(\widehat{\alpha} N+\frac{\widehat{\beta}}{N}\right)+\widehat{\delta}\frac{N(N+1)}{N-1}\right].
\end{equation}
\item
P6: {\bf{for $\beta=1/2$ we have}}
\begin{eqnarray}
f_1&=&-1-\frac{1}{3}\left(\frac{1}{N}+\frac{1}{N-1}+\frac{1}{N-t}\right),\quad
f_2=\frac{1}{\sqrt{6\pi}}\left(\frac{1}{t}+\frac{1}{t-1}+\frac{1}{N-t}\right),\\
f_3&=&-\frac{N(N-1)(N-t)}{4\pi t^2 (t-1)^2}\left[\widehat{\alpha}+\widehat{\beta}\frac{t}{N^2}+\widehat{\gamma}\frac{t-1}{(N-1)^2}+\widehat{\delta}\frac{t(t-1)}{(N-t)^2}\right],
\end{eqnarray}
{\bf{while for $\beta=1$ we find}}
\begin{equation}
\widetilde{f}_1+\widetilde{f}_2=-1-\frac{1}{3}\left(\frac{1}{N}+\frac{1}{N-1}+\frac{1}{N-t}\right),
\end{equation}
\begin{equation}
\widetilde{f}_3=\frac{1}{4\pi}\left(\frac{1}{t}+\frac{1}{t-1}+\frac{1}{N-t}\right)\dot{N}-\frac{N(N-1)(N-t)}{4\pi t^2 (t-1)^2}\left[\widehat{\alpha}+\widehat{\beta}\frac{t}{N^2}+\widehat{\gamma}\frac{t-1}{(N-1)^2}+\widehat{\delta}\frac{t(t-1)}{(N-t)^2}\right].
\end{equation}
\end{itemize}
\end{itemize}
Equation (\ref{master}) allows to construct further FLRW cosmological models not covered in the work of \cite{Myrzakulov,Dark}. For instance, by setting $k=0$ in (\ref{master}) and choosing
\begin{equation}
f_1=-1,\quad f_2=0,\quad
f_3=\frac{1}{4\pi}\left(\frac{\gamma}{t}+\frac{\delta}{t-1}+\frac{\epsilon}{t-a}\right)\dot{N}
+\frac{\alpha\beta t-q}{4\pi t(t-1)(t-a)}N
\end{equation}
with $\alpha,\beta,\gamma,\delta,\epsilon,a,q\in\mathbb{R}$ such that $\epsilon=\alpha+\beta-\gamma-\delta+1$ ensures that the regular singularity at infinity has exponents $\alpha$ and $\beta$, one may derive a Heun cosmology. For a detailed study  on the Heun functions we refer to \cite{Ronveaux}. It is interesting to observe that equation (\ref{master}) permits to construct a FLRW cosmological model such that $N$, i.e. the logarithm of the scale factor, satisfies a certain Gambier equation \cite{Gam}. {\bf{For instance, if we let
\begin{equation}
k=0,\quad\beta=\frac{1}{2},\quad
f_1=-\frac{2+3N}{N},\quad
f_2=-\frac{1}{\sqrt{6\pi}N},\quad
f_3=-\frac{1}{4\pi}\left[\xi(t)N^2-N\frac{d}{dt}\left(\frac{\dot{\xi}}{\xi}\right)\right]
\end{equation}
in (\ref{master}) where $\xi$ is some differentiable function, then (\ref{master}) reduces to the Gambier G15 equation (\ref{GXV}).}}
The canonical list of second-order ordinary differential equations with the Painlev\'e property remains till date a rich area of research and contains certain outstanding problems. The minimal list of twenty four equations given by Gambier are fundamental in the sense that the remaining ones can be obtained through Miura transformations \cite{LGR}. Even though the Gambier equations  were classified more than a hundred
years ago \cite{Gam,Ince} they have led a somewhat forgotten existence
compared to the Painlev\'{e} equations which have received a great deal of attention not
only because of their inherent structures but also on account of
their crucial role in defining the very notion of integrability
 of nonlinear partial differential equations appearing in
various branches of the physical sciences. Among the equations of Gambier's list some  are linearizable, and equation G27
when  written as a system has the form
\be\label{G1}\dot{y}=-y^2+by+c, \;\;\dot{x}=a_g x^2+nxy+s\ee
where $a_g,b$ and $c$ are functions of the independent variable $t$,
$s$ is a constant and $n$ is an integer. Eqn. (\ref{G1}) is generally referred to as the Gambier equation
and may be expressed as a
second-order equation \cite{GGCG}   
\be\label{G2}
\ddot{x}=\left(1-\frac{1}{n}\right)\frac{\dot{x}^2}{x}
+a_g\frac{n+2}{n} x\dot{x}
+b\dot{x}
-s\left(1-\frac{2}{n}\right)\frac{\dot{x}}{x}
-\frac{a_g^2}{n}x^3+(\dot{a}_g-a_g b)x^2+\left(cn-\frac{2a_g s}{n}\right)x-bs
-\frac{s^2}{nx}.
\ee 
It was shown in \cite{GGCG} how almost all the linearizable equations G1-G26 of the Gambier list can be obtained from G27 by taking appropriate limits.

\section{Reductions of the $2+1$-FLRW cosmology model to the Gambier equations}
In \cite{WK} Williams and Kevrekidis examined the reduction of models of the  FLRW type in $2 + 1$ dimensions and analysed the  resulting ODEs. In the following, we briefly outline their derivation of the Ermakov-Pinney equation. The Einstein's equation for the metric given by  
  \be\label{met2}ds^2=-dt^2+a^2(t)\left(\frac{dr^2}{1-kr^2}+r^2 d\varphi^2\right),\ee
  are
\begin{align}\label{WK1}\frac{\dot{a}^2+k}{a^2}&=\widehat{G}(\rho_m +\rho_\phi),\\
\frac{\ddot{a}}{a}&=-\widehat{G}(p_m+p_\phi),\\
\ddot{\phi}+2\frac{\dot{a}}{a}\dot{\phi}&=-\frac{\partial
V}{\partial \phi}.\end{align} 
Here, $\widehat{G}$ is the Newton gravitational constant in $2+1$-dimensions while $\rho$ and $p$ represent the density and pressure, respectively. The subscript $m$ refers to the matter (fluid) and $\phi$ to the homogeneous scalar field whose  density and pressure are given by $\rho_\phi=\dot{\phi}^2/2+V(\phi)$ and $p_\phi=\dot{\phi}^2/2-V(\phi)$, respectively, and the scalar field $\phi$ can be viewed as a kind of perfect fluid with equation of state $p_\phi=w_\phi\rho_\phi$ where
\begin{equation}
w_\phi=\frac{\dot{\phi}^2-2V(\phi)}{\dot{\phi}^2+2V(\phi)}.
\end{equation}
Furthermore, by assuming an equation of state for the matter $p_m=w_m\rho_m$ with $0<w_m\le 1$, the authors in \cite{WK} derived the following expression for the matter density
\be\label{dichte}
\rho_m=\xi a^{-2(1+w_m)}, 
\ee
where $\xi=\rho_m(0)a(0)^{2(1+w_m)}$ is a constant. From (\ref{WK1}) and the forms of $\rho_\phi$ and $p_\phi$
stated above one can derive the equation
\be\label{WK2}\frac{\ddot{a}}{a}-\frac{\dot{a}^2}{a^2}=-\widehat{G}\left[\dot{\phi}^2+\frac{\xi(1+w_m)}{a^{2(1+w_m)}} \right]+\frac{k}{a^2}.\ee
 Introducing a new temporal variable $\tau=\tau(t)$ and $Y=Y(\tau)$ such that
\be\label{conditions}
Y(\tau(t))=a^{(1+w_m)}\quad\mbox{with}\quad \frac{dt}{d\tau}=\frac{1}{Y(\tau)}
\ee 
 we find that  (\ref{WK2}) may be expressed in the form
 \be\label{WK3}
 Y^{\prime\prime}(\tau)+
 \widehat{G}(1+w_m)(\phi^{'})^2Y=-\frac{\widehat{G}\xi(1+w_m)^2}{Y^3}+\frac{k(1+w_m)}{Y^{1+\frac{2}{1+w_m}}}.\ee
with $^{'}=d/d\tau$. In the case of flat universes ($k=0$), this equation reduces to the
Ermakov-Pinney equation with potential $Q(\tau)=\widehat{G}(1+w_m)(\phi^{'})^2$ according to the notation used in \cite{WK,HL}. The authors of \cite{WK} then go on to derive explicit solutions of the
original cosmological models by assuming a solution of
(\ref{WK3}). In particular, they consider the special case when $\xi=0$ (corresponding to either a vanishing initial density or a vanishing initial scale factor, the latter provided that $w_m<-1$) and
employ a specific ansatz for the potential term in order to derive
some solutions of (\ref{WK3}) under the condition that $\rho_m(0)=0$. At this point a remark is in order. Such an initial constraint coupled with (\ref{dichte}) implies that at all times there is no matter content in the cosmological models developed by \cite{WK}. Moreover, the equation of state $p_m=w_m\rho_m$ leads to a vanishing pressure. This is equivalent to have a vanishing energy-momentum tensor for the matter component and therefore, the choice of taking $\gamma=1$ (or equivalently $w_m=0$), i.e. a cold dust scenario, in the models studied by \cite{WK} is shaky from a physics-based perspective.

While most  papers on  FLRW cosmology exploit the Ermakov-Pinney equation, our motivation here is to show that
(\ref{WK3}) is mappable to certain reductions of the Gambier equation. According to existing observational data, it is not yet possible to distinguish between phantom ($w_m<-1$) and non-phantom cosmological scenarios. Hence, it is physically reasonable to consider a phantom energy case with $w_m<-1$. This usually leads to a Big Rip. In fact, in the reminder of this paper we relax the previous restrictions on the parameter $\gamma=w_m+1$ \cite{WK} and allow it to be even negative, i.e. $w_m<0$. Then, under the transformation $Y\rightarrow X=Y^\sigma$ with $\sigma\in\mathbb{R}$,  (\ref{WK3}) becomes
 \be\label{WK4}
X^{\prime\prime}=\left(1-\frac{1}{\sigma}\right)\frac{{X^{'}}^2}{X}-\widehat{G}\sigma(1+w_m)(\phi^{'})^2 X-\widehat{G}\xi\sigma(1+w_m)^2 X^{1-\frac{4}{\sigma}}+k\sigma(1+w_m)X^{1-\frac{2}{\sigma}\left(1+\frac{1}{1+w_m}\right)}.\ee
If $k\neq 0$, we can set $\sigma = - 4/k$ and $1+w_m = - k/2l$ with $l\neq 0$,  and (\ref{WK4})
assumes the following form
 \be\label{WK4a}
X^{\prime\prime}=\left(1+\frac{k}{4}\right)\frac{{X^{'}}^2}{X}-\frac{2\widehat{G}}{l}(\phi^{'})^2X+\widehat{G}\xi\frac{k}{l^2}X^{1+ k}+\frac{2k}{l}X^{1+\frac{k}{2}-l},\ee
which  is a reduction of the Gambier equation. It is worth to note that this reduced Gambier equation
is obtained from the $2+1$-FLRW cosmology model as proposed by \cite{WK}.  Since it follows
from the cosmology equation so the signs of the coefficients are fixed by physics.

\subsection{Generalized Pinney Equation}
Let us set $\lambda(\tau)= \widetilde{G}(d\phi/d\tau)^2$ and $\mu= \widetilde{G}\xi=2\pi G\rho_m(0)(a(0))^{2(1+w_m)}$ in (\ref{WK4a})
 and introduce the transformation $X=(z/v)^{1/\alpha}$ with $\alpha\neq 0$ where
 $v=$constant. Under such a transformation (\ref{WK4a}) reduces to
 \be\label{WK4b}
 z^{\prime\prime}=\left(1+\frac{k}{4\alpha}\right)\frac{\dot{z}^2}{z}
 -\frac{2\alpha}{l}\lambda(\tau)z+\frac{\alpha k}{l^2}\mu
 v^{-\frac{k}{\alpha}}z^{1+\frac{k}{\alpha}}+\frac{2\alpha
 k}{l}v^{-\frac{1}{\alpha}\left(\frac{k}{2}-l\right)}z^{1+\frac{1}{\alpha}\left(\frac{k}{2}-l\right)}.\ee 
Demanding that the coefficient of $\dot{z}^2$ vanishes leads to $\alpha=-k/4=\pm 1/4$ and setting 
\be\label{omega}
\omega(\tau)=\frac{\lambda(\tau)}{2}=\frac{1}{2}\widetilde{G}(\phi^{'})^2
\ee
allows to cast (\ref{WK4b}) into the form
\be\label{WK4c}z^{\prime\prime}=\frac{k}{l}\omega(\tau) z-\frac{\mu
k^2}{4l^2}\frac{v^4}{z^3}-\frac{k^2}{2l}v^{2-4l/k}z^{-1+4l/k}.\ee
Finally, writing $n=-l/k$  gives rise to
\be\label{WK4d}z^{\prime\prime}+\frac{1}{n}\omega(\tau) z=-\frac{\mu
v^4}{4n^2}\frac{1}{z^3}+\frac{k}{2n}v^{2(2n+1)}z^{-(4n+1)}.\ee We now
distinguish the following cases.
\begin{itemize}
\item
{\underline{\bf Case 1:}}  Let $n=1$ which in view of $\gamma=-k/2l$ implies that $\gamma=1/2$ or equivalently $w_m=-1/2$ and causes  (\ref{WK4d}) to appear as
\be z^{\prime\prime}+\omega(\tau) z=-\frac{\mu v^4}{4z^3}-\frac{k v^6}{2z^5}.\ee

\item
{\underline{\bf Case 2:}} Let $n=1/2$ so that $\gamma=1$, i.e. $w_m=0$. For this we find that (\ref{WK4d}) becomes
\be 
z^{\prime\prime}+2\omega(\tau) z=\frac{(k-\mu)v^4}{z^3}.
\ee 
As a consistency check, we observe that for $\gamma=1$ the pressure vanishes, i.e. $p_m=0$, which is in line with what we  expect for a cold dust scenario in a $(2+1)$-FLRW cosmological model. It is interesting to observe that in the case $\phi(\tau)$ is a constant scalar field, i.e. the potential $V$ is constant, then $\omega(\tau)=0$ in virtue of (\ref{omega}) and the above equation admits the solution
\be\label{zeta}
z(\tau)=\sqrt{c_1\tau^2+2c_1 c_2\tau+c_1c_2^2+\frac{(k-\mu)v^4}{c_1}}
\ee
with $c_1$ and $c_2$ arbitrary integration constants and $c_1\neq 0$. The corresponding solution with a minus sign in front of the root has been neglected because it would lead to a negative scale factor. Since $\alpha=-k/4$ simple algebra gives that $Y=z/v$ and by construction it is a solution to (\ref{WK3}) for $k=-1$ (open universe) or $k=1$ (closed universe). At this point a remark is in order. It is gratifying to observe that despite the assertion in \cite{WK} according to which equation (\ref{WK3}) cannot be solved explicitly, we were able to provide a nontrivial solution. If we recall that $w_m=0$, the first equation in (\ref{conditions}) implies that $Y=a$ and we have
\be\label{scaletta}
a(\tau)=\frac{1}{v}\sqrt{c_1\tau^2+2c_1 c_2\tau+c_1c_2^2+\frac{(k-\mu)v^4}{c_1}}.
\ee
In order to switch back to the time variable $t$, we need to integrate the ODE
\be
\frac{dt}{d\tau}=\frac{v}{z(\tau)}.
\ee
Let us consider for simplicity the case $c_2=0$ and $c_1>0$ and break down the analysis as follows
\begin{itemize}
\item
{\underline{\bf Case $k=-1$:}} Then, we have
\be
\frac{dt}{d\tau}=\frac{v}{\sqrt{c_1}\sqrt{\tau^2-A}},\quad A=\frac{(1+\mu)v^4}{c_1^2}>0.
\ee
A trivial integration gives 
\be
t=\frac{v}{\sqrt{c_1}}\ln{(\tau+\sqrt{\tau^2-A})}.
\ee
Since $v$ is a free parameter, we can exploit this degree of freedom in order to choose $v=\sqrt{c_1}$ and we end up with
\be
t=\ln{(\tau+\sqrt{\tau^2-(1+\mu)})}.
\ee
Solving for $\tau$ gives 
\begin{equation}
\tau=\frac{e^{2t}+1+\mu}{2e^t}
\end{equation}
and by means of (\ref{scaletta}) the scale factor reads 
\be
a(t)=\sqrt{\left(\frac{e^{2t}+1+\mu}{2e^t}\right)^2-(1+\mu)}.
\ee
It can be immediately seen that such a universe will collapse at the time $t_c=\ln{\sqrt{1+\mu}}$. 
\item
{\underline{\bf Case $k=+1$:}} Fixing $v=\sqrt{c_1}$ yields
\be
\frac{dt}{d\tau}=\frac{1}{\sqrt{\tau^2+1-\mu}}.
\ee
If $\mu<1$, a trivial integration gives $t=\sinh^{-1}{(\tau/\sqrt{1-\mu})}$ and hence, $\tau=\sqrt{1-\mu}\sinh{t}$. The scale factor is represented by
\be
a(t)=\sqrt{1-\mu}\cosh{t}.
\ee
In the special case $\mu=1$, we get $a(t)=e^t$.

\end{itemize}
\item
{\underline{\bf Case 3:}} Let $n=3/8$. Then, $\gamma=4/3$ or equivalently, $w_m=1/3$. This corresponds to the case of relativistic matter and (\ref{WK4d}) reads
\be
z^{\prime\prime}+\frac{8}{3}\omega(\tau) z=-\frac{16\mu v^4}{9z^3}+\frac{4kv^{7/2}}{3z^{5/2}}.
\ee

\item
{\underline{\bf Case 4:}} Let $n=1/3$ (so that $\gamma=3/2$) whence   (\ref{WK4d}) becomes
\be z^{\prime\prime}+3\omega(\tau) z=-\frac{9\mu v^4}{4z^3}-+\frac{3k v^{10/3}}{2z^{7/3}}.\ee
However, as $\gamma=3/2$ one has $p_m=\rho_m/2$ and this obviously conforms to the original assumption made in \cite{WK}, namely $1\le \gamma<2$ which implies $1/4\le n<1/2$.

\item
{\underline{\bf Case 5:}} Let $n=1/4$ (so that $\gamma=2$ or equivalently $w_m=1$) whence  (\ref{WK4d}) becomes
\be z^{\prime\prime}+4\omega(\tau) z=-\frac{4\mu v^4}{z^3}+\frac{2kv^3}{z^2}.\ee
This corresponds to $p_m=\rho_m$, i.e. a stiff equation of state.\\

\item
{\underline{\bf Case 6:}}(Ermakov-Painlev\'e II equation)\, Let $n = -1$ which implies $\gamma = -1/2$ and set $\delta=k/2v^2$. Then, (\ref{WK4d}) becomes
\be
z^{\prime\prime}-\omega(\tau)z + \delta z^3 = -\frac{1}{4z^3}\left(\frac{\sqrt{\mu}k}{2\delta}\right)^2.
\ee
This nonlinear equation is somewhat similar in appearance  to the (single component) Ermakov-Painlev\'e II equation
\be\label{EPII} 
{z}^{\prime \prime}+\frac{\tau}{2}z+\delta z^3=-\frac{1}{4z^3}\left(\nu-\frac{\delta}{2}\right)^2
\ee
which has been derived by \cite{Conte,Rogers}. \cite{Gromak} showed that (\ref{EPII}) is related   to the
Painlev\'e II equation. Furthermore, \cite{Rogers} proved that the canonical single component Ermakov-Painlev\'e II equation
can be related to a particular Ermakov-Ray-Reid system ( for details, see \cite{Rogers}) whose characteristic invariant may be exploited systematically to construct the solutions. We will not further elaborate on this aspect because we leave it for future investigation.
\end{itemize}
Finally, it is worth noting that all the generalized Pinney equations as obtained in this section follow
from  the cosmological Gambier equation (\ref{WK4a}) or (\ref{WK4b}) where the signs of the coefficients are fixed by the underlying physical problem.

\section{The Milne-Pinney system and the generalized FLRW cosmology equation with time dependent drag term}
In the previous section, we have mapped the reduced $2+1$-FLRW cosmology equation (\ref{WK4a}) to various equations of the  Pinney type. In the present section, we address instead the question of what happens when the cosmological equation (\ref{WK4a}) is modified by adding a time-dependent drag term. This can happen when the scalar field exhibits a time-dependent behavior. In this case, even though it is not possible to map the generalized FLRW cosmology equation to 
the regular Gambier equation G27, it turns out that it can be mapped to a generalization of the parametric Gambier equation G27 (see appendix). When the Pinney cosmology equation is modified with a time-dependent drag term the resulting
second-order nonlinear differential equation becomes 
\be\label{x3}
\ddot{q} + w(t)^2\dot{q} + Kq^{-3} = 0, \qquad \dot{q} = \frac{dq}{dt}, 
\ee 
and describes the time
evolution of a damped oscillator with inverse quadratic potential. This is
the Milne-Pinney equation and in this context, it may  be regarded as the  Milne-Pinney cosmology equation. It is possible
once again to map the Milne-Pinney equation to the Gambier equation. We list here below the most important cases under the assumption that $q$ satisfies the Milne-Pinney equation (\ref{x3}).
\begin{itemize}
\item
{\underline{\bf Case I:}} Let $x = a(t)/q^2(t)$. Then, $x$ satisfies the Gambier equation
\be \label{pGam}\ddot{x} =
\frac{3}{2}\frac{{\dot{x}}^2}{x} -
\frac{\dot{a}}{a}\dot{x} + \frac{2K}{a^2}x^3 +
\left[\frac{\ddot{a}}{a} - \frac{1}{2}\left(\frac{\dot{a}}{a}\right)^2 +2w^2(t)\right]x. \ee
\item
{\underline{\bf Case II:}} Let $x = a(t)/q^4(t)$. Then, $x$ satisfies the Gambier equation
\be \label{pGam1}\ddot{x} =
\frac{5}{4}\frac{{\dot{x}}^2}{x} -\frac{1}{2}
\frac{\dot{a}}{a}\dot{x} + \frac{4K}{a}x^2 +
\left[\frac{\ddot{a}}{a} - \frac{3}{4}\left(\frac{\dot{a}}{a}\right)^2 +4w^2(t)\right]x. \ee
\item
{\underline{\bf Case III:}} Let $x = a(t)q^4(t)$. Then, $x$ satisfies the Gambier equation
\be \label{pGam2}\ddot{x} =
\frac{3}{4}\frac{{\dot{x}}^2}{x} +\frac{1}{2}
\frac{\dot{a}}{a}\dot{x} +
\left[\frac{\ddot{a}}{a} - \frac{5}{4}\left(\frac{\dot{a}}{a}\right)^2 -4w^2(t)\right]x -4Ka. \ee
\item
{\underline{\bf Case IV:}} Let $x = a(t)q^2(t)$. Then, $x$ satisfies the Gambier equation
\be \label{pGam3}\ddot{x} =
\frac{1}{2}\frac{{\dot{x}}^2}{x} +
\frac{\dot{a}}{a}\dot{x} +
\left[\frac{\ddot{a}}{a} - \frac{3}{2}\left(\frac{\dot{a}}{a}\right)^2 -2w^2(t)\right]x -\frac{2Ka^2}{x}. \ee
\end{itemize}
It is clear that the equations (\ref{WK4a}) and (\ref{pGam})-(\ref{pGam3}) are the
reductions of the parametric equation (\ref{E.9}) drived in the appendix. In other words, the
generalized FLRW equation is not the reduction of the Gambier equation but of its parametric generalization.

\section{Conclusions}

In this work, we have shown how the Gambier equations emerge in $3+1$- and $2+1$-FLRW cosmologies. This was done by deriving a master equation for the logarithm of the scale factor under the assumption of an EOS generalizing a family of EOS widely used to describe DM \cite{Lopez,N1,N2}. Our findings reveals the importance of future studies devoted to construct analytical/approximated solutions of the Gambier equations in order to better understand the behaviour of Gambier inspired cosmological models. Even though, the role of the Ermakov-Pinney equation in FLRW cosmology is well known in the literature, here we have explored the role of various generalizations of Pinney equation, as formulated by \cite{Rogers,Rogers1,RSW}, in 2+1-dimensional FLRW cosmology. This raises an important issue regarding the existence of  a systematic relation between integrable class of 1+1-ODEs and FLRW cosmology and clearly, demands further investigation.

\section{Appendix: The generalized parametric Gambier equation}
 Let us consider the following Riccati differential
 operator 
 \be\label{E.7}
 \R=\frac{d}{dt}+H(x,t),\quad H(x,t)=h(t)x+c(t)+\frac{\sigma}{x},
 \ee
 with $\sigma$ being a constant. Then, $\R x=0$ implies that
 \be\label{E.8}
 \dot{x}+H(x,t)x=0,
 \ee
 and requiring that $\R^2 x=0$ yields
 \begin{equation}
 \R(\dot{x}+h(t)x^2+c(t)x+\sigma)=0
 \end{equation}
 which can be written as
\be\label{E.9}
\ddot{x}+3h(t)x\dot{x}+2c(t)\dot{x}+\sigma\frac{\dot{x}}{x}+
 h^2(t)x^3+\left[2h(t)c(t)+\dot{h}\right]x^2+\left[c^2(t)+\dot{c}+2\sigma
 h(t)\right]x+2\sigma c(t)+\frac{\sigma^2}{x}=0.
 \ee
  This equation bears a remarkable similarity with the Gambier
  equation, which is given by
$$\ddot{x}=\frac{n-1}{n}\frac{\dot{x}^2}{x}+\alpha(t)\frac{n+2}{n}x\dot{x}+\beta(t)\dot{x}-
\frac{n-2}{n}\frac{\dot{x}}{x}\sigma(t)+\frac{\alpha^2(t)}{n}x^3+$$
\be\label{E.10}\left(\dot{\alpha}
-\alpha(t)\beta(t)\right)x^2+\left[\gamma(t)
n-2\alpha(t)\frac{\sigma(t)}{n}\right]x-\beta(t)\sigma(t)-\frac{\sigma^2(t)}{n
x}.\ee For $n=1$ under the scaling transformation
$t\longrightarrow -t$ this equations becomes \be\label{E.11}
\ddot{x}+3\alpha(t)x\dot{x}+\beta(t)\dot{x}+\frac{\dot{x}}{x}\sigma(t)+\alpha^2(t)x^3+
(\alpha^\prime(t)+\alpha\beta)x^2+(2\alpha
\sigma(t)-\gamma(t))x+\beta\sigma(t)+\frac{\sigma^2(t)}{x}=0.\ee

Clearly if we assume $\sigma$ to be independent of $t$ in
(\ref{E.11}) then its comparison with (\ref{E.9}) shows that
$$h(t)=\alpha(t),\;\;\;2c(t)=\beta(t)$$ while $c(t)$ must satisfy
the equation \be\label{E.12}c^\prime(t)+c^2(t)+\gamma(t)=0.\ee
Eqn. (\ref{E.12}) is the generalized Riccati-I equation and under
the transformation $c(t)=\frac{u^\prime(t)}{u}$ becomes
\be\label{E.13} u^{\prime\prime}+\gamma(t)u=0,\ee which is a
linear equation. In fact if $\gamma(t)>0$ for all $t$ then it
becomes a harmonic oscillator equation with variable frequency. In
general  (\ref{E.10}) also contains a term involving
$\dot{x}^2/x$. Such a term may also be included into the previous
sequence of equations if we make the following observation. Define
an operator
\be\label{E.14}\T=\left(\frac{d}{dt}+\frac{\dot{x}}{x}\right)\ee
and consider the following sequence of differential equations
$\T^m x=0$:
$$\T x=2\dot{x}=0, \;\;\T^2
x=\left(\frac{d}{dt}+\frac{\dot{x}}{x}\right)2\dot{x}=2\ddot{x}+2\frac{\dot{x}^2}{x}=0,\;\mbox{etc.}$$
This suggests that we  consider a combination of the two operators
$\R$ and $\T$, namely $\R+\lambda \T$, where $\lambda$ is a
parameter. Then we have
$$(\R+\lambda \T)x=\dot{x}+h(t)x^2+c(t)x+\sigma +2\lambda
\dot{x}=0,$$ or in other words
$$(1+2\lambda)\dot{x}+h(t)x^2+c(t)x+\sigma=0,$$ which is once
again a generalized Riccati-I equation. Next consider the equation
$$(\R^2+\lambda \T^2)x=0$$ which implies
$$(1+2\lambda)\ddot{x}+2\lambda
\frac{\dot{x}^2}{x}+3h(t)x\dot{x}+2c(t)\dot{x}+\sigma\frac{\dot{x}}{x}+h^2(t)x^3$$
$$+\left(h^\prime(t)2h(t)c(t)\right)x^2+\left(c^\prime(t)+c^2(t)+2\sigma
h(t) \right)x+2\sigma c(t)+\frac{\sigma^2}{x}=0.$$

\section*{Acknowledgement}
PG is profoundly grateful to Professors Basil Grammaticos and Alfred Ramani
for discussing various things regarding the Gambier XXVII equation.
Work by the author PG was supported by the Khalifa University of 
Science and Technology under grant number FSU-2021-014. Last but not least, we would like to thank the anonymous referee for his/her valuable suggestions which definitely improved the present work.

\bigskip

{\bf Data accessibility } This article does not use data.

\end{document}